\documentclass[prl,twocolumn,showpacs]{revtex4}

\usepackage{graphicx}
\usepackage{amsmath}
\usepackage{epstopdf}
\usepackage{amsmath}
\usepackage{amsfonts}

\usepackage{epsfig} 
\usepackage{color} 

\newcommand{\Ket}[1]{\left|#1  \right>}

\newcommand{\Braket}[1]{\left<#1  \right>}

\begin{document}

\title{Edge Physics of the Quantum Spin Hall Insulator from a Quantum Dot Excited by Optical Absorption}

\author{Romain Vasseur and Joel E. Moore}

\affiliation{Department of Physics, University of California, Berkeley, Berkeley CA 94720, USA}
\affiliation{Materials Science Division, Lawrence Berkeley National Laboratory, Berkeley CA 94720, USA}

\date{\today}

\begin{abstract}

The gapless edge modes of the Quantum Spin Hall insulator form a helical liquid in which the direction of motion along the edge is determined by the spin orientation of the electrons. In order to probe the Luttinger liquid physics of these edge states and their interaction with a magnetic (Kondo) impurity, we consider a setup where the helical liquid is tunnel-coupled to a semiconductor quantum dot which is excited by optical absorption, thereby inducing an effective quantum quench of the tunneling. At low energy, the absorption spectrum is dominated by a power-law singularity. The corresponding exponent is directly related to the interaction strength (Luttinger parameter) and can be computed exactly using boundary conformal field theory thanks to the unique nature of the Quantum Spin Hall edge.

%The finite temperature crossover is obtained exactly in that regime. We discuss how our proposal could be used to measure experimentally the Luttinger parameter in a more controlled way than using transport properties.

\end{abstract}


\maketitle

Recent experiments on optical absorption of a quantum dot coupled to a metallic reservoir offer a new window into the correlated electron state that underlies the Kondo effect~\cite{QuenchKondo1,QuenchKondo2}. The absorption of a single photon can be treated as a sudden change of the Hamiltonian (a ``quench''): the exchange interaction between the dot and reservoir is abruptly changed by the absorption.  Over a range of (shifted)  photon energies between the observation temperature $T$ and Kondo temperature $T_K$, a power-law in the absorption spectrum is observed.   This power law is a consequence of the Anderson orthogonality catastrophe~\cite{andersonorthogonality}: the overlap between two metallic states differing by the presence or absence of a scattering potential goes to zero algebraically in the number of electrons $N$.  It is natural to ask whether other kinds of metals that also support Kondo effects can be probed with this type of experiment, and what such an experiment would reveal.

The main goal of this Letter is to explain the optical absorption of a few-electron quantum dot in the Kondo regime when it is coupled to the helical metal of electrons at the edge of a quantum spin Hall effect (QSHE) droplet.  The QSHE edge~\cite{KaleMele1,KaleMele2} is a one-dimensional metal where spin plays a fundamental role: as a result of spin-orbit coupling, there is a single time-reversal-related ``Kramers pair'' of low-energy propagating modes, which can be pictured as a right-moving mode of electrons with spin up along some axis and a left-moving mode with spin down.  This edge is robust to disorder and interactions as long as the original symmetries of time-reversal and charge are unbroken~\cite{QSHedge1,QSHedge2}.  When time-reversal is broken by a static magnetic perturbation, the conductance goes to zero and can be computed via integrability of point tunneling in a Luttinger liquid~\cite{BarrierQSHE}. The Anderson impurity problem, which is our starting point, is more complex as the impurity is dynamical and time-reversal is preserved.

QSHE edges have been probed through transport on (Hg,Cd)Te quantum wells~\cite{HgTe,HgTeExp}, InAs/GaSb quantum wells~\cite{rrdu}, and (in the former material) through SQUID imaging of the generated magnetic flux~\cite{moler}.  The Kondo effect along the QSHE edge has previously been studied theoretically for its effects on transport~\cite{QSHKondo1,QSHKondo2,altshuleryudson} (see also {\it e.g.}~\cite{Transport0,Transport1,Transport2,Transport3} for other studies involving transport), where it has little effect at least for a single impurity: the DC conductance remains $2 e^2 / h$, just as in the absence of the impurity.  This might suggest that Kondo effects in the QSHE are subtle and difficult to observe.  To the contrary, the Kondo impurity's effects on optical absorption are much clearer than in transport: again a power-law is observed in the absorption spectrum, but now the power-law is determined by the interaction strength (Luttinger parameter) along the edge because special properties of the QSHE edge fix a scaling dimension in the field theory of this problem.

Unlike an ordinary Luttinger liquid~\cite{KondoLuttinger}, the helical liquid plus Kondo impurity can be mapped exactly onto the Kondo problem in an ordinary Fermi liquid, where the interactions in the QSHE edge generate an effective anisotropic exchange coupling.  In fact a different version of this mapping was noted~\cite{KondoSchiller} before the helical liquid was understood to arise in physical systems. This remarkable property of the QSHE edge enables us to solve exactly the absorption problem at low energy.  In the same regime realized in existing experiments with Fermi liquids, we find that the absorption spectrum shows a power-law tail with an exponent given by the Luttinger parameter, thus providing a direct measurement of interaction strength in helical liquids.

\paragraph{Physical setup}

We start with the description of the Quantum Spin Hall edge as a helical liquid (HL) with counterpropagating modes with opposite spins along some axis, with forward scattering interactions~\cite{QSHedge1,QSHedge2} 
\begin{equation}
{\cal H}_{\rm HL}(x)=\Psi^\dagger \left( - i v_F \sigma_z \partial_x \right) \Psi + g \psi^\dag_{\uparrow} \psi_{\uparrow} \psi^\dag_{\downarrow} \psi_{\downarrow},
\end{equation}
with $\Psi(x) = \left(\psi_\uparrow \ \psi_\downarrow \right)^T$ a two-dimensional spinor. We consider interactions at the edge only and ignore bulk interactions that would lead to more exotic behaviors~\cite{Bulk1, Bulk2}. Single-particle backscattering terms that would open a gap in the Luttinger liquid are not allowed by time-reversal symmetry. Following~\cite{QuenchKondo0,QuenchKondo1}, this HL is tunnel-coupled to a semiconductor quantum dot (QD) whose charge state is controlled by an external gate voltage $V_g$. This gate voltage can be tuned such that the topmost occupied level $h$ (``valence level'') lies far below the Fermi energy, so that it can be considered as occupied. On the other hand, the conduction level $d$ can be considered as unoccupied initially. We then apply a circularly polarized light beam (say with polarization $\uparrow$) with frequency $\omega$. This will excite an electron with spin $\uparrow$ from the valence level $h$ into the conduction level $d$, thus leaving a positively charged hole behind with spin $\downarrow$ in the $h$ level (see~\cite{QuenchKondo1,QuenchKondo2,QuenchKondo3} for a related protocol in the case of a dot coupled to a Fermi reservoir). This effectively induces an attractive Coulomb interaction $U_{eh}$ between the excited electron and the hole -- this can also be thought of as the energy difference between the state with both $d$ and $h$ levels occupied (with Coulomb repulsion between the electrons), and the state with only the $d$ level occupied. This hole is assumed to be stable and static compared to the other time scales of the problem. Let $d^\dagger_\sigma$ and $h^\dagger_\sigma$ be the electron/hole creation operators in the $d$ and $h$ levels, respectively. The quantum dot/light interaction is thus described by $H_{\rm L} \propto \left(d^\dagger_{\uparrow} h^\dagger_{\downarrow} \mathrm{e}^{- i \omega t} + {\rm h.c.} \right)$. 

\begin{figure}[t!]
\includegraphics[width=1.1\linewidth]{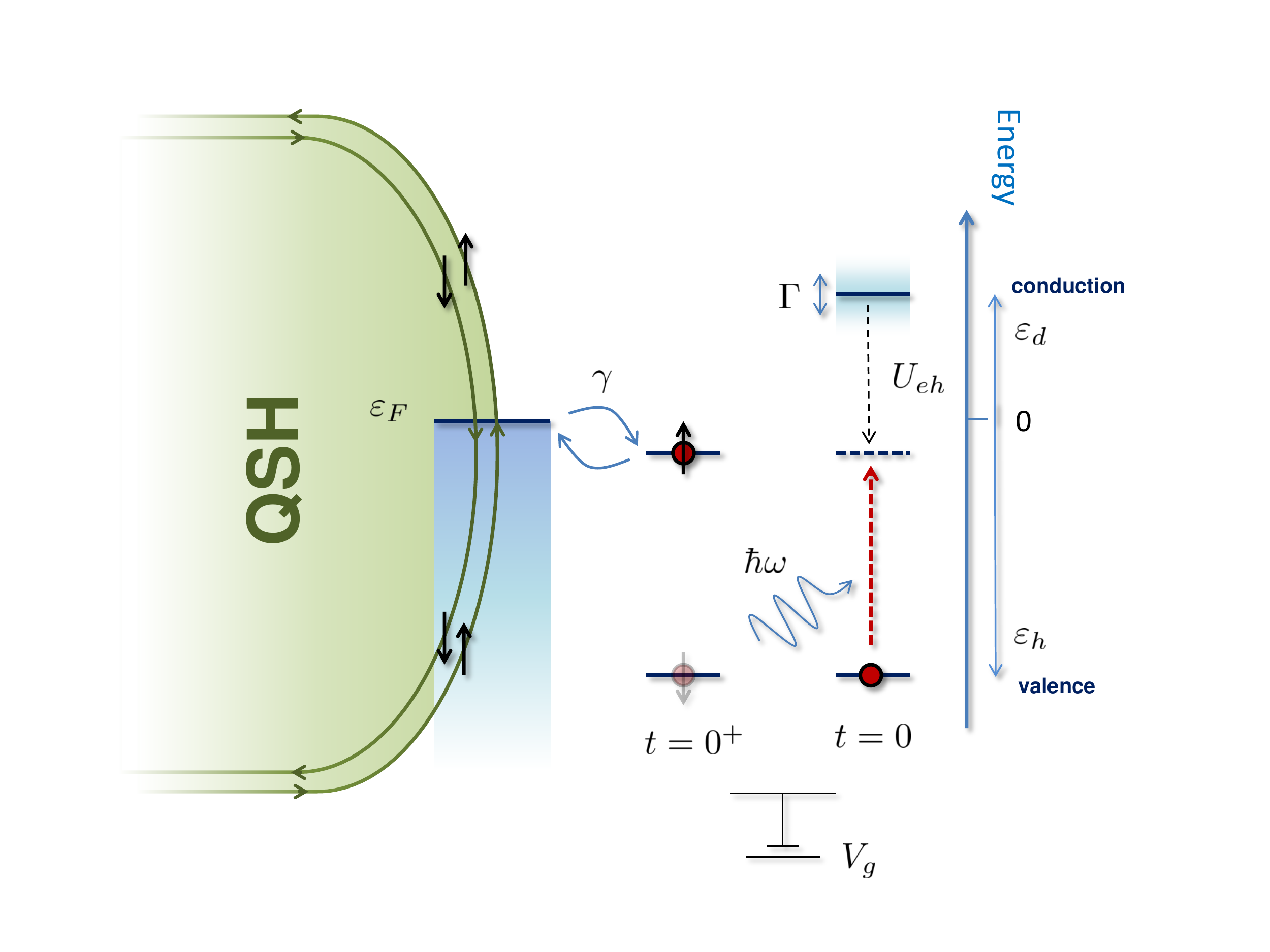}
\caption{Physical setup: the photon absorption of a quantum dot implements a quantum quench of the tunneling between the dot and a helical liquid. The Luttinger liquid properties of the helical liquid can be read off from the low energy part of the absorption spectrum. }
\label{fig1}
\end{figure}

\paragraph{Anderson impurity model.}
The relevant part of the Hamiltonian of the quantum dot is then given by
%
%\begin{equation}
$
H_{\rm dot} = U  n^d_\uparrow  n^d_\downarrow + \epsilon_d n^d + \epsilon_h n^h -  U_{eh} n^d n^h, 
$
%\end{equation}
%
with $n^d_\sigma = d^\dagger_\sigma d_\sigma$, $n^h_\sigma = h^\dagger_\sigma h_\sigma$, and $n^{d/h} = \sum_\sigma n^{d/h}_\sigma$. All the energies are measured with respect to the Fermi energy $\epsilon_F=0$ and are positive. 
The energy cost for creating a hole in the $h$ level $ \epsilon_h$ is assumed to be very large (compared to the bulk gap), so the $h$ level can be considered as occupied initially -- {\it i.e.}, before the light perturbation is turned on. The hole degree of freedom will be integrated out in the following but it is important to keep track of it in order to trigger the quantum quench.
Finally, the tunneling between the HL and the QD reads
\begin{equation}
H_{\rm t} = \gamma \sum_{\sigma} \left( d^\dagger_\sigma \psi_\sigma(0) + {\rm h.c.} \right).
\end{equation}
This gives a (bare) hybridization width $\Gamma = \gamma^2 \pi \rho_0$ to the $d$ level, where $\rho_0$ is the density of states (per spin) of the helical liquid.  
%Note that this parameter will be renormalized by the interactions in the helical liquid. 
In the following, we will neglect the two-particle backscattering by the impurity. The full Hamiltonian is  $H=H_{\rm HL}+H_{\rm dot}+ H_{\rm t}$ where $H_{\rm HL}=\int dx {\cal H}_{\rm HL}$. We remark that even if the two modes of the helical edge are not fully spin polarized, the coupling to the impurity in the Anderson Hamiltonian will have the same form~\cite{SuppMat}. 

We have assumed that $\epsilon_d$ does not depend on the spin of the $d$ electrons, which can be achieved by adding a term that accounts for an applied bias on the HL. Let us hence assign a different chemical potential $\pm V/2$ to the right- and left-moving electrons  $H_{\rm V} = -\frac{V}{2} \int dx \left(\psi^\dag_{\uparrow} \psi_{\uparrow} -\psi^\dag_{\downarrow} \psi_{\downarrow} \right)$. Because the total number of electrons with a given spin is conserved by the Hamiltonian, this term can be replaced by an effective magnetic field acting on the dot, thereby inducing an effective Zeeman splitting $\epsilon^{\uparrow\downarrow}_d = \epsilon_d \pm \frac{V}{2}$. 
%We remark that this trivial effect of the bias in a HL coupled to an impurity is precisely the reason why the correction due to the impurity to the conductance $G=2e^2/h$ vanishes in the dc limit~\cite{QSHKondo1,QSHKondo2}. 
In the following, we will assume that this bias term is tuned such that $\epsilon^{\uparrow}_d = \epsilon^{\downarrow}_d = \epsilon_d$. A genuine magnetic field (except along the $z$ axis) would open a gap in the HL.

\paragraph{Absorption spectrum and Loschmidt echo.}  

We remark that the Hamiltonian conserves $n_\sigma^h$, so we can easily integrate out the hole degree of freedom. Using Fermi's golden rule, the absorption spectrum of the photons can be expressed as 
\begin{equation}
A(\omega) = \kappa \sum_{m,l} \rho^i_m \left| ~_f\langle l \right| d^\dagger_{\uparrow} \left| m \rangle_i \right|^2 \delta \left( \omega - (E^f_l-E^i_m)\right), 
\end{equation}
where $\kappa$ is a proportionality constant that will depend on the precise experimental setup. The labels $\alpha=i,f$ correspond to the initial and final Hamiltonians 
\begin{equation}
H^{\alpha} = H_{\rm HL}+ H_{\rm t} + U  n^d_\uparrow  n^d_\downarrow + \epsilon_d n^d + \delta_{\alpha,f}\left(\epsilon_h - U_{eh} n^d \right), 
\end{equation}
with $H^{\alpha} \Ket{n}_\alpha=E_n^{\alpha}  \Ket{n}_\alpha$, and $\rho^i_m=\Braket{m  | \rho^i | m}=\frac{1}{Z} \mathrm{e}^{-E^i_m/T}$ is the initial thermal density matrix of the system. With the parameters described previously (in particular, $\epsilon_d, U \gg \Gamma$), the $d$ level can initially be considered as completely empty and decoupled from the helical liquid, and the absorption of the photon induces an effective quantum quench of the tunneling between the dot and the helical liquid. We can thus write the eigenstates of $H^i$ as product states over the helical liquid and the unoccupied dot $\Ket{m}_i \simeq \Ket{m}_{\rm HL} \otimes \Ket{0}_d$. The quantity $P(\omega)=\kappa^{-1} A(\omega)$ then corresponds to the distribution of the work done during a quantum quench~\cite{Silva,WorkSpectra} starting from a decoupled HL at temperature $T$ (supposed to be smaller than all the other energy scales) and a QD in the state $\Ket{\uparrow}_d$, with the quench corresponding to suddenly turning on the tunneling between the helical liquid and the dot (see Fig.~\ref{fig1}). The Fourier transform of the work distribution is known~\cite{Silva} as the Loschmidt echo
\begin{equation}
G(t) = \langle \mathrm{e}^{i H_i t} \mathrm{e}^{-i H_f t}  \rangle_i 
\label{eqLoschmidt}
\end{equation}
where $ \langle \dots  \rangle_i$ refers to a thermal average over the initial density matrix $\rho_i = \frac{\mathrm{e}^{-H_i/T}}{Z_i}$, with the dot in the state $\Ket{\uparrow}_d$. We shall show in the following that the large time behavior of this Loschmidt echo can be computed exactly.

\paragraph{From Anderson to Kondo.}

Because we are dealing with a strongly interacting many-body problem, the real-time dynamics of the system after the quantum quench is extremely complicated, and is controlled by several different energy scales. In what follows, we will restrict ourselves initially to the particle-hole symmetric case $\epsilon_d-U_{eh} = -\frac{U}{2}$ of the Anderson impurity model, and consider the case $\Gamma \lesssim U/2$. We will also shift the frequencies by $\omega_0=E^f_0-E^i_0$, the groundstate energy difference between $H_i$ and $H_f$, which corresponds to the minimal work needed at zero temperature to perform the quantum quench ($\omega_0 \sim \epsilon_h -U/2$). Let us assume that we are in a frequency regime such that $T, \nu=\omega-\omega_0 \ll U$, so that one can effectively integrate out the charge degrees of freedom on the dot to go from the Anderson problem to a reduced Kondo setup~\cite{Kondo} --  although the initial quantum dot setup is crucial to trigger physically the quantum quench using optical absorption. The resulting effective Hamiltonian density at low energy is~\cite{QSHKondo1,QSHKondo2}
\begin{equation}
{\cal H}^f_{\rm eff} = {\cal H}_{\rm HL} + \delta(x) J \vec{S}.\left(\Psi^\dagger \frac{\vec{\sigma}}{2}\Psi \right),  
\end{equation}
with $\vec{S} = \sum_{\sigma \sigma'} d^\dagger_\sigma \frac{\vec{\tau}_{\sigma \sigma'}}{2} d_{\sigma'}$ the local spin on the dot.  (There is no potential scattering term induced at the symmetric point.) In terms of the Anderson model parameters, the bare Kondo coupling reads $J=\frac{8 \gamma^2}{U}$ at the particle-hole symmetric point. This tunneling term should be considered energy-dependent and will be renormalized by the interactions in the HL. We also point out that the coupling to the effective Kondo impurity remains isotropic even in the presence of spin-orbit coupling~\cite{Isotropic} (see also~\cite{QSHKondo1,QSHKondo2,SuppMat}).

\paragraph{Bosonization.} 

To proceed, we bosonize the HL electrons by introducing $\psi_{\uparrow \downarrow} = \frac{1}{\sqrt{2 \pi}} \mathrm{e}^{\pm i \sqrt{4 \pi} \phi_{\uparrow \downarrow}}$. Using standard bosonization formulas, one obtains a (spinless) Luttinger liquid Hamiltonian for the HL $H_{\rm HL} = \frac{v}{2}  \int dx \left[ \Pi^2+ (\partial_x \Phi)^2 \right] $ where we have introduced the non-chiral boson $\Phi = (\phi_{\uparrow}+\phi_{\downarrow})/\sqrt{K}$, and its dual $\theta = \sqrt{K}(\phi_{\uparrow}-\phi_{\downarrow})$ with $\Pi=-\partial_x \theta$. The renormalized velocity reads $v = \sqrt{v_F^2 - (g/2\pi)^2}$, and the Luttinger liquid parameter is given by $K = \sqrt{\frac{v_F - \frac{g}{2 \pi}}{v_F + \frac{g}{2 \pi}}} \ < \ 1$. Using standard canonical transformations and taking the scaling limit, the bosonized form of the Kondo interaction is $\frac{J}{4 \pi} \left(\mathrm{e}^{-i \sqrt{4 \pi K} \Phi(0)} S^{+} + {\rm h.c.} \right)$, mixing right and left-movers $\Phi = \phi_R + \phi_L$.  This Hamiltonian is purely chiral in terms of the left-moving fields $\varphi^{\rm e/o}(x)=\left(\phi_L(x) \pm \phi_R(-x) \right)/\sqrt{2}$~\cite{EdgeCFT,FQHEtunnel}. The odd boson then decouples, while the chiral Hamiltonian for $\varphi \equiv \varphi^{\rm e}$ reads
\begin{equation}
 H^f_{\rm eff} = \int d x (\partial_x \varphi)^2  + \frac{J}{4 \pi} \left(\mathrm{e}^{-i \sqrt{8 \pi K} \varphi(0)} S^{+} + {\rm h.c.} \right).
 \label{eqKondoQSH}
\end{equation}
This is the bosonized form of the one-channel anisotropic Kondo effect in a Fermi liquid. Put differently, the Kondo problem in a helical liquid can be mapped onto the usual Kondo effect in a Fermi liquid, but the Coulomb interactions in the HL induce an effective anisotropy in the Kondo coupling after this transformation~\cite{KondoSchiller}. This mapping is due to the unique nature of the HL, and does not apply to the case of a Kondo impurity in the usual spinful Luttinger liquid~\cite{KondoLuttinger}. It is worth pointing out at this point that up to another canonical transformation, eq.~\eqref{eqKondoQSH} is related to the Interacting Resonant Level Model, a problem that has attracted a lot of attention recently due to the development of exact methods out of equilibrium~\cite{IRLM1,IRLM2,IRLM3}.

\begin{figure}[t!]
\includegraphics[width=1.1\linewidth]{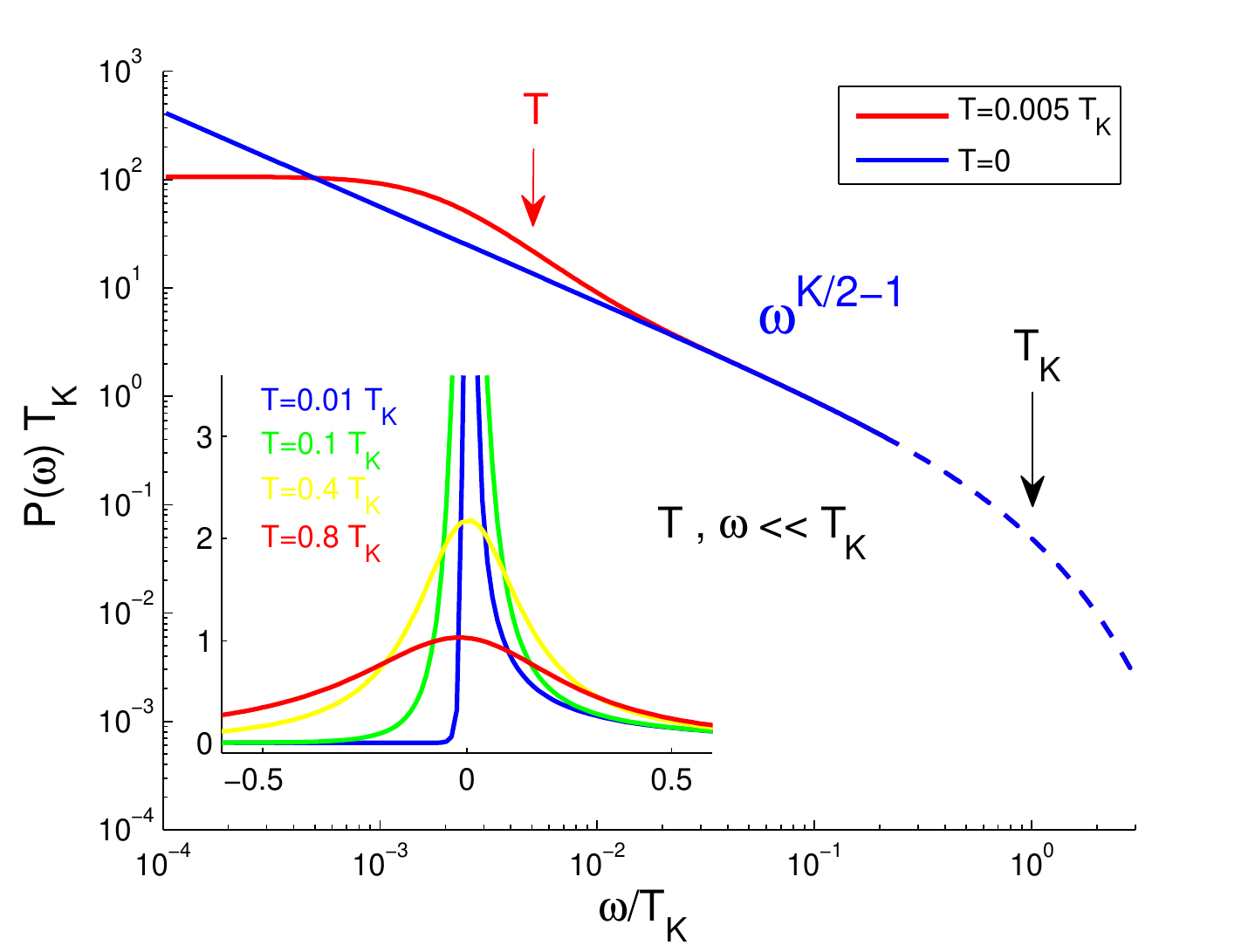}
\caption{Absorption spectrum (work distribution) in the Kondo regime $T,\nu=\omega-\omega_0 \ll T_K$ computed from BCFT (with $K=1/4$ in this example). Inset: finite-temperature smearing of the edge singularity. }
\label{fig2}
\end{figure}

\paragraph{Low energy Kondo physics.}

The impurity perturbation has dimension $h=K$, so it is always relevant for repulsive Coulomb interactions. At lowest-order, the Kondo temperature reads $T_K \propto J^{1/(1-K)}$. A more precise form of the Kondo temperature in our problem is given by~\cite{KondoSchiller,QSHKondo1}
\begin{equation}
T_K = D \exp \left(-\frac{ \pi U}{8 \Gamma} \frac{{\rm arcsinh} \xi }{\xi}\right),
\end{equation}
with $D$ the bandwidth and $\xi = \sqrt{\left(1+\frac{\pi U}{8 \Gamma}(1-K) \right)^2-1 }$ the Coulomb-induced anisotropy parameter. We expect this formula to be valid only at low energy compared to the dot Coulomb interaction $U$. In the absence of interaction in the QSH helical liquid ($K=1$), one recovers the well-known form of the Kondo temperature in the Anderson/Kondo problem.  We will focus on the low-temperature regime $T, \nu \ll T_K$ that is appropriate for the existing experiments with Fermi liquid leads.  In the crossover regime where $T, \nu$ are comparable to $T_K$ but still less than $U$, numerical renormalization group methods might be useful to obtain the intermediate behavior.

\paragraph{Zero temperature edge singularity.}

In the zero temperature limit $T=0$, there is no energy scale remaining in the problem so we can tackle this low energy (large time $t \gg T_K^{-1}$) limit using Boundary Conformal Field Theory (BCFT). Folding the chiral anisotropic Kondo Hamiltonian~\eqref{eqKondoQSH}, the Kondo impurity then becomes a boundary condition on a non-chiral boson. At low energy, this flows to a conformally invariant boundary condition and the Loschmidt echo can be argued to be related to the two-point function of a boundary condition changing operator (BCC)~\cite{QuenchBCC} (see also~\cite{LoschmidtCFT}). This operator $\varphi\equiv\varphi_{i \to f}$ has dimension $ h_{\rm BCC} = \frac{K}{4} $ so we recover the $\delta_{\uparrow \downarrow}=\pm \frac{\pi}{2}$ phase shifts of the electrons in the usual case of noninteracting leads ($K=1$) as $\frac{1}{4} = \frac{1}{2} \left(\frac{\delta_\uparrow}{\pi} \right)^2 + \frac{1}{2} \left(\frac{\delta_\downarrow}{\pi} \right)^2$~\cite{KondoKM,EdgeCFT}, while the value for $K=\frac{1}{2}$ can be interpreted as twice the dimension of the spin operator in the Ising model~\cite{QuenchBCC}. The Loschmidt echo thus behaves as $G(t) \sim t^{-K/2}$ at large time. In terms of the absorption spectrum, this means that we expect the following edge singularity
\begin{equation}
A(\omega) \underset{\omega-\omega_0 \ll T_K}{\propto} \frac{1}{T_K}\theta(\omega-\omega_0)\left( \frac{\omega-\omega_0}{T_K}\right)^{K/2-1}.
\label{eqEdgesingularity}
\end{equation}
The low energy part of the absorption spectrum thus contains a clear signature of the Luttinger physics of the HL, very different from what one would obtain in the case of an ordinary (spinful) Luttinger liquid~\cite{KondoLuttinger}. For $K=1$, one recovers the $(\omega-\omega_0)^{-1/2}$ behavior observed experimentally for a Fermi liquid reservoir~\cite{QuenchKondo2}. 
 
\paragraph{Finite temperature crossover.}

At non-zero temperature $T \ll T_K$, the Loschmidt echo can still be thought of as a two-point function of BCC operators. Indeed, introducing $\varphi^\dagger = \varphi_{f \to i}$ with the normalization $\varphi^\dagger \varphi =1$, one can recast~\eqref{eqLoschmidt} as $G(t) = \langle \varphi^\dagger(t) \varphi(0) \rangle_i $, where the time evolution is performed with the initial Hamiltonian $H_i$. This finite temperature two-point function can readily be computed using a conformal mapping
\begin{equation}
G(t) \underset{t \gg T_K^{-1}}{\propto} \left(\frac{\pi T}{\sin i \pi T t} \right)^{-K/2}, 
\end{equation}
with $T_K$ playing the role of a UV cutoff. The absorption spectrum can then be obtained by Fourier transform (see Fig.~\ref{fig2}). The edge singularity~\eqref{eqEdgesingularity} is smeared at finite temperature, so it is important to work in the regime $T \ll \nu \ll T_K$ to have the power-law behavior~\eqref{eqEdgesingularity}. However, we emphasize that as long as $T \ll T_K$, the full finite temperature crossover is captured by CFT, providing a wider frequency range to measure the exponent $K$. In that low energy regime, the finite temperature absorption spectrum could also be used to test non-equilibrium fluctuation theorems~\cite{Crooks}. 

\paragraph{Discussion.} 

Our proposal to measure the interaction strength along the QSHE edge using optical absorption requires one to work in the Kondo regime $\nu=\omega-\omega_0,T \ll T_K \ll U$. For estimated (Hg,Cd)Te QD and HL parameters $U \sim 10$ meV, $\Gamma \sim 1$ meV and $D \sim 10$ meV, one finds $T_K \sim 2 - 10$ K depending on the Luttinger parameter $K$, with larger values of $T_K$ corresponding to strong Coulomb interactions. The Luttinger liquid nature of the helical liquid thus makes it easier to reach the Kondo regime (large $T_K$ with $T_K \ll U$), which was already accessed experimentally for a Fermi liquid reservoir~\cite{QuenchKondo2}.  For the temperatures used in the (Hg,Cd)Te quantum well experiment~\cite{HgTeExp}, one expects to have $T/T_K \sim 0.01 - 0.003$ so that the low energy BCFT results should hold.   To our knowledge few-electron quantum dots have not yet been created in this material, so it may be more practical to use InAs/GaSb quantum wells~\cite{rrdu}, in which the QSHE also survives to higher temperatures.  Although we have derived eq.~\eqref{eqEdgesingularity} at the particle-hole symmetric point $\epsilon_d-U_{eh} = -\frac{U}{2}$, we expect this formula to remain valid whenever the dot is approximately occupied by one electron only~\cite{QuenchKondo1}, in part because potential scattering terms $\psi(0)_\uparrow^\dagger \psi(0)_\downarrow+h.c.$ that would change the exponent in eq.~\eqref{eqEdgesingularity} are not allowed by time-reversal symmetry.

Our results suggest that optical absorption could be a reasonable alternative to transport in order to probe the edge physics of topological phases of matter. It would be interesting to see whether similar quantum dot setups could provide new insights on the physics of topological Kondo systems~\cite{TopoKondo}, or help in probing Majorana modes at the edge of topological superconductors~\cite{Interplay}.

\smallskip

\paragraph{Acknowledgments.}
This work was supported by the Quantum Materials program of LBNL (R.V) and NSF DMR-1206515 (J.E.M.). We thank J.~Dahlhaus and D.~Goldhaber-Gordon for insightful discussions. RV also wishes to thank D. Kennes, V. Meden and H. Saleur for collaborations on related matters.

\newpage

\onecolumngrid

\centerline{\bf Supplementary material}

\bigskip

\section{Anderson Hamiltonian for a non fully spin-polarized helical liquid}

A natural question about the Hamiltonian used in the main text is whether errors might be induced by writing the quantum spin Hall edge as spin-polarized along some direction, with right-moving electrons having spin up and left-moving electrons having spin down.  It is often emphasized that the quantum spin Hall edge survives even when no spin direction is conserved, and that the edge should be viewed as a ``Kramers pair'' of states conjugate under the time-reversal operation.  This supplement reviews what it means to say that the edge does not have a well-defined spin direction, and why in principle that could lead to a reduction in the tunneling amplitude $\gamma$, although the Hamiltonian will still have the same form.

In practice, the degree of spin polarization at the quantum spin Hall edge is thought to be high, and $\gamma$ is usually taken as a measured quantity rather than computed theoretically, but the considerations here may be useful for other problems.  First, it is simple to show that the form of the Hamiltonian remains correct even if the edge is not made up of spin eigenstates.  Consider the action of the tunneling Hamiltonian that takes the low-lying right-moving state into some superposition in the two-level system made of the spin states of one electron on the dot:
\begin{equation}
H_R = \psi_R(0) (\gamma_1 d^\dagger_\uparrow + \gamma_2 d^\dagger_\downarrow) + {\rm h.c.}
\end{equation}
The action of the tunneling Hamiltonian on the left-moving state is then obtained by acting with the time-reversal operation:
\begin{equation}
H_L = \psi_L(0) (-\gamma^*_2 d^\dagger_\uparrow + \gamma^*_1 d^\dagger_{\downarrow}) + {\rm h.c.}
\end{equation}
where we have used the convention that the time-reversal operator is $\Theta = i \sigma^y K$, with $K$ denoting complex conjugation.  Then an $SU(2)$ unitary transformation of spin basis on the dot
\begin{equation}
\label{eqChangeBasis}
\left( \begin{matrix}
{\tilde d}^\dagger_\uparrow \\
{\tilde d}^\dagger_\downarrow
\end{matrix} \right) = {1 \over \gamma} \left( \begin{matrix}
\gamma_1 & \gamma_2 \\
-\gamma_2^* & \gamma_1^*
\end{matrix} \right) \left( \begin{matrix}
{d}^\dagger_\uparrow \\
{d}^\dagger_\downarrow
\end{matrix} \right).
\end{equation}
gives the form assumed in the main text, with $\gamma = \sqrt{|\gamma_1|^2 + |\gamma_2|^2}$:
\begin{equation}
H_{\rm tun} = H_L + H_R = \gamma \psi_R(0) {\tilde d}^\dagger_\uparrow + \gamma \psi_L(0) {\tilde d}^\dagger_\downarrow + {\rm h.c.}
\end{equation}
The tunneling term thus remains the same. It is also easy to see that even though the interaction term on the dot $U n_{\uparrow}n_{\downarrow} $ is not invariant under the transformation~\eqref{eqChangeBasis}, the additional terms can be readily absorbed by a redefinition of $\epsilon_d$. Therefore, the full Hamiltonian has exactly the same form if the helical edge is not fully spin-polarized, albeit with (slightly) renormalized coefficients. In particular, the coupling to the effective magnetic impurity in the Kondo limit remains isotropic even in the presence of spin-orbit coupling.  A similar conclusion was reached in~\cite{Isotropic} in a different language.  

\section{Physical picture of the reduced-spin-polarization phenomenon}

This result seems perhaps counterintuitive if the edge is not fully spin-polarized: the dot with one electron is a two-level system and hence can be interpreted as a maximally polarized spin-half state along some direction.  To give a physical picture of what is the ``rotated'' spin basis on the dot when the edge state is not a spin eigenstate to start with, it may be simplest to work with a simple example of a non-spin-polarized Bloch state and think about a photoemission or tunneling experiment.  Consider for simplicity the following (not yet normalized) Bloch state of crystal momentum zero, obtained by superposing a spin-up Gaussian centered on locations $x = n a$, $n \in \mathbb{Z}$, with a spin-down Gaussian cented on sites $x = (n + 1/2) a$:
\begin{equation}
|\psi\rangle = \sum_{n = -\infty}^\infty \left( {1 \over \sqrt{2 \pi \sigma^2}} e^{-(x - n a)^2 / (2 \sigma^2)} |\uparrow\rangle + {1 \over \sqrt{2 \pi \sigma^2}} e^{-(x - n a - a/2)^2 / (2 \sigma^2)} |\downarrow\rangle \right).
\end{equation}
As $\sigma \rightarrow \infty$, this becomes the uniform state with $k=0$ and spin state $|\uparrow\rangle$ + $|\downarrow \rangle$, i.e., perfectly polarized along the ${\bf +\hat x}$ spin direction.  However, at finite $\sigma$ the expansion of the above wavefunction over plane waves (momentum eigenstates), as would be detected in an idealized photoemission experiment, becomes more interesting: the expansion of the above wavefunction in plane waves is
\begin{equation}
|\psi \rangle = C \sum_{n = -\infty}^{n = \infty} e^{- \sigma^2 n^2/(2 a^2)} e^{ i n x / a} \left(|\uparrow \rangle + (-1)^n |\downarrow \rangle \right),
\end{equation}
where $C$ is a normalization constant.  We see that the plane-wave expansion now includes all momenta differing from the crystal momentum (zero) by a reciprocal lattice vector, and that plane waves with odd $n$ are spin-polarized along ${\bf -\hat x}$.  The absolute normalization can be set by letting $C=1$ at $\sigma = \infty$ (a single plane wave appears).  Then, $|\psi \rangle$ will be consistently normalized for all $\sigma$ if
\begin{equation}
C^{-2} = \sum_{n = -\infty}^{n = \infty} e^{- \sigma^2 n^2 / (a^2)} = \theta_3(0,e^{-\sigma^2/a^2}).
\end{equation}
Now the meaning of the statement that this state is not spin-polarized is evident: while $|\psi\rangle$ is a pure quantum state, it involves multiple momenta, and consequently a measurement that only probes spin will give probabilities consistent with a mixed spin state rather than a pure spin state.  (In modern parlance the state shows ``entanglement'' between the spin and momentum Hilbert spaces.)  The average spin along the $x$ direction is
\begin{equation}
C^2 \sum_{n = - \infty}^{n = \infty} (-1)^n e^{-n^2 \sigma^2/a^2} = { \theta_3(0,e^{-4 \sigma^2/a^2}) - \theta_2 (0,e^{-4 \sigma^2/a^2}) \over \theta_3 (0, e^{-\sigma^2/a^2})},
\end{equation}
which drops from 1 at $\sigma = \infty$, to 0.17 at $\sigma = a$,  to zero at $\sigma = 0$.  The average of a measurement in the other spin directions is always zero, proving that (except at $\sigma=\infty$ where only a single momentum state is involved) the spin is not fully polarized if momentum is not resolved.

The tunneling problem above can now be analyzed similarly and shows the same effect.  To start with, assume that the fundamental tunneling is spin-independent: then the magnitude of the effective tunneling strength $\gamma$ to the dot can be reduced if the different momenta contained in the Bloch state interfere destructively because they tunnel into the dot with opposite spin directions.  If the tunneling has its own spin-orbit coupling, then the details of this interference will be modified depending on the specific momentum dependence of the spin direction induced in the tunneling.  To summarize, the incomplete spin polarization in a Bloch state can reduce the total tunneling matrix element $\gamma$, but the form of the Hamiltonian in the main text remains valid, and the reduction of $\gamma$ is only expected to be large if the spin polarization of the edge state is small, which is not believed to be the case for (Hg,Cd)Te.


\begin{thebibliography}{99}

\bibitem{QuenchKondo1} H. E. T\"ureci {\it et al.}, 
%{\sl  Many-Body Dynamics of Exciton Creation in a Quantum Dot by Optical Absorption: A Quantum Quench towards Kondo Correlations}, 
Phys. Rev. Lett. {\bf 106}, 107402 (2011).

\bibitem{QuenchKondo2}  C. Latta {\it et al.},
%{\sl Quantum quench of Kondo correlations in optical absorption},
Nature {\bf 474}, 627--630 (2011).

\bibitem{andersonorthogonality} P.W. Anderson, Phys. Rev. Lett. {\bf 18}, 1049--1051 (1967).

\bibitem{KaleMele1} C.L. Kane and E.J. Mele, Phys. Rev. Lett. {\bf 95}, 226801 (2005).

\bibitem{KaleMele2} C.L. Kane and E.J. Mele, Phys. Rev. Lett. {\bf 95}, 146802 (2005).

\bibitem{QSHedge1} C. Wu, B.A. Bernevig and S-C. Zhang,  
%{\sl Helical Liquid and the Edge of Quantum Spin Hall Systems}, 
Phys. Rev. Lett. {\bf 96}, 106401 (2006).

\bibitem{QSHedge2} C. Xu and J.E. Moore,   
Phys. Rev. B {\bf 73}, 045322 (2006).

\bibitem{BarrierQSHE} R. Ilan, J. Cayssol, J.H. Bardarson and J.E. Moore,  Phys. Rev. Lett. {\bf 109}, 216602 (2012).

\bibitem{HgTe} B. A. Bernevig, T. L. Hughes, and S. C. Zhang, Science {\bf 314}, 1757 (2006).

\bibitem{HgTeExp} M. K\"onig {\it et al.}, Science {\bf 318}, 766 (2007).

\bibitem{rrdu} I. Knez, R.-R. Du, and G. Sullivan, Phys. Rev. Lett. {\bf 107}, 136603 (2011).

\bibitem{moler} K.C. Nowack {\it et al.}, Nature Materials {\bf 12}, 787--791 (2013).


\bibitem{QSHKondo1} J. Maciejko, C. Liu, Y. Oreg, X-L. Qi, C. Wu and S-C. Zhang, 
%{\sl Kondo Effect in the Helical Edge Liquid of the Quantum Spin Hall State}, 
Phys. Rev. Lett. {\bf 102}, 256803 (2009).

\bibitem{QSHKondo2} Y. Tanaka, A. Furusaki and K.A. Matveev, 
%{\sl Conductance of a Helical Edge Liquid Coupled to a Magnetic Impurity}, 
Phys. Rev. Lett. {\bf 106}, 236402 (2011).

\bibitem{altshuleryudson} B.L. Altshuler, I.L. Aleiner and V.I. Yudson, Phys. Rev. Lett {\bf 111}, 086401 (2013).

\bibitem{Transport0} C.-Y. Hou, E.-A. Kim and C. Chamon, Phys. Rev. Lett. {\bf 102}, 076602 (2009). 
%Corner Junction as a Probe of Helical Edge States

\bibitem{Transport1} G. Dolcetto, S. Barbarino, D. Ferraro, N. Magnoli and M. Sassetti, Phys. Rev. B {\bf 85}, 195138 (2012).
%Tunneling between helical edge states through extended contacts

\bibitem{Transport2}  J.-R. Souquet and P. Simon, Phys. Rev. B {\bf 86}, 161410(R) (2012).
%Finite frequency noise in a quantum point contact between helical edge states

\bibitem{Transport3} Y.-L. Lee and Y.-W. Lee, Phys. Rev. B {\bf 88}, 035112 (2013).
%Electrically tunable two-channel Kondo fixed points in helical liquids

\bibitem{KondoLuttinger} D-H. Lee and J. Toner, Phys. Rev. Lett. {\bf 69}, 3378--3381 (1992).
%{\sl Kondo effect in a Luttinger liquid },

\bibitem{KondoSchiller} A. Schiller and K. Ingersent , 
%{\sl Exact results for the Kondo effect in a Luttinger liquid}, 
Phys. Rev. B {\bf 51}, 4676--4679 (1995).


\bibitem{Bulk1} M. Hohenadler, T. C. Lang, and F. F. Assaad, Phys. Rev. Lett. {\bf 106}, 100403 (2011).

\bibitem{Bulk2} E.-G. Moon, C. Xu, Y.B. Kim and L. Balents, Phys. Rev. Lett. {\bf 111}, 206401 (2013).




\bibitem{QuenchKondo0} R.W. Helmes, M. Sindel, L. Borda and J. von Delft, Phys. Rev. B {\bf 72}, 125301 (2005).


\bibitem{QuenchKondo3} M. Heyl and S. Kehrein, 
%{\sl  X-ray edge singularity in optical spectra of quantum dots}, 
Phys. Rev. B {\bf 85}, 155413 (2012).

\bibitem{Silva} A. Silva, 
%{\sl Statistics of the Work Done on a Quantum Critical System by Quenching a Control Parameter}, 
Phys. Rev. Lett. {\bf  101}, 120603 (2008).


\bibitem{WorkSpectra} M. Heyl and S. Kehrein,  
%{\sl Crooks Relation in Optical Spectra: Universality in Work Distributions for Weak Local Quenches}, 
Phys. Rev. Lett. {\bf 108}, 190601 (2012).

\bibitem{SuppMat} See supplemental material.

%\bibitem{SchriefferWolff} J.R. Schrieffer and P. A. Wolff, Phys. Rev. {\bf 149}, 491 (1966).

\bibitem{Kondo}  A. Hewson, {\sl The Kondo Problem to Heavy Fermions, Cambridge Studies in Magnetism} (Cambridge University Press, Cambridge, England, 1997).

\bibitem{Isotropic} L. Shekhtman, O. Entin-Wohlman and A. Aharony, Phys. Rev. Lett. {\bf 69}, 836 (1992).

\bibitem{EdgeCFT} I. Affleck and A.W.W. Ludwig, J. Phys. A: Math. Gen. {\bf 27}, 5375 (1994). 

\bibitem{FQHEtunnel} P. Fendley, A.W.W. Ludwig and H. Saleur, Phys. Rev. Lett. {\bf 74}, 3005 (1995).



\bibitem{IRLM1} B. Doyon, Phys. Rev. Lett. {\bf 99}, 076806 (2007).

\bibitem{IRLM2} E. Boulat and H. Saleur, Phys. Rev. B {\bf 77}, 033409 (2008).

\bibitem{IRLM3} E. Boulat, H. Saleur and P. Schmitteckert, Phys. Rev. Lett. {\bf 101}, 140601 (2008).

\bibitem{QuenchBCC} R. Vasseur, K. Trinh, S. Haas and H. Saleur, 
%{\sl Crossover Physics in the Nonequilibrium Dynamics of Quenched Quantum Impurity Systems}, 
Phys. Rev. Lett. {\bf 110}, 240601 (2013).

\bibitem{LoschmidtCFT} J-M. St\'ephan and J. Dubail, J. Stat. Mech. (2011) P08019.

\bibitem{KondoKM} I. Affleck, Nucl. Phys. B {\bf 336}, 517--532 (1990).

\bibitem{Crooks} M. Heyl and S. Kehrein, Phys. Rev. Lett. {\bf 108}, 190601 (2012).

\bibitem{TopoKondo} B. B\'eri and N.R. Cooper, Phys. Rev. Lett. {\bf 109}, 156803 (2012). 

\bibitem{Interplay} M. Cheng, M. Becker, B. Bauer and R.M. Lutchyn, {\em Interplay between Kondo and Majorana interactions in quantum dots}, {\tt arXiv:1308.4156}.







\end{thebibliography}
\end{document}